\documentclass[dvips]{arxstspdf}
\usepackage{flushend}
\usepackage{stfloats}

\volume{24}
\issue{3}
\pubyear{2009}
\firstpage{343}
\lastpage{360}
\doi{10.1214/09-STS301}

\makeatletter
\newproclaim{Remark}{Remark}
\makeatother

\begin{document}
\begin{frontmatter}

\title{The Impact of Levene's Test of Equality of Variances on Statistical
Theory and Practice}
\runtitle{The Impact of Levene's Test of Equality}

\begin{aug}
\author[a]{\fnms{Joseph L.} \snm{Gastwirth}\corref{}\ead[label=e1]{jlgast@gwu.edu}},
\author[b]{\fnms{Yulia R.} \snm{Gel}\ead[label=e2]{ygl@math.uwaterloo.ca}}
\and
\author[c]{\fnms{Weiwen} \snm{Miao}\ead[label=e3]{wmiao@haverford.edu}}
\runauthor{J. L. Gastwirth, Y. R. Gel and W. Miao}

\affiliation{George Washington University, University of Waterloo, Canada, and Haverford College}

\address[a]{Joseph L. Gastwirth is Professor of Statistics and Economics,
George Washington University,
Pennsylvania Ave. 2140, Washington, DC   20052, USA \printead{e1}.}
\address[b]{Yulia R. Gel is Associate Professor, Department of Mathematics
and Statistics, University of Waterloo, 200 University Ave. W, Waterloo,
Ontario, Canada   N2L 3G1  \printead{e2}.}
\address[c]{Weiwen Miao is Associate Professor, Department of Mathematics,
Haverford College, Haverford, Pennsylvania  19041, USA \printead{e3}.}

\end{aug}

\begin{abstract}
In many applications, the underlying scientific question concerns
whether the variances of $k$ samples are equal. There are
a substantial number of tests for this problem. Many of them rely on the
assumption of normality and are not robust to its violation. In
1960 Professor Howard Levene proposed a new approach to this problem by applying the
$F$-test to the absolute deviations of the observations from their
group means. Levene's approach is powerful and robust to
nonnormality  and  became a very popular tool for checking the
homogeneity of variances.

This paper reviews the original method proposed by Levene and subsequent robust
modifications. A modification of Levene-type tests to increase their power
to detect monotonic trends in variances is discussed. This procedure is useful when one is
concerned with an alternative of increasing or decreasing
variability, for example, increasing volatility of stocks prices or ``open or
closed gramophones'' in regression residual analysis. A major section
of the paper is devoted to discussion of various scientific problems
where Levene-type tests have been used, for example, economic anthropology,
accuracy of medical measurements, volatility of the price of oil,
studies of the consistency of jury awards in legal cases and the
effect of hurricanes on ecological systems.
\end{abstract}


\end{frontmatter}

\section*{Introduction}

Very few statisticians write an article that is still cited forty or
fifty years after it is published. Professor Howard Levene, whose
research focused on statistical problems arising in biological
science, was the sole author of three such classic papers. Not only
have they been cited hundreds of times; they continue to be cited
today. Professor Levene passed away in July, 2003 and this article is
written in recognition of his important contributions to statistical
science.

After introducing two earlier well cited articles, Levene (\citeyear{1949Levene}) and
Levene (\citeyear{1953Levene}), the impact of the third article, on a robust test for
the equality of the variances of $k$ populations, will be emphasized.
In particular, both the robustness aspect and the focus
on the ``spread'' or variability of the data in the Levene (\citeyear{1960Levene})
article influenced the work of the authors, especially J.~L. Gastwirth,
who took his first class in Mathematical Statistics from Professor
Levene.

The first seminal article of Professor Levene concerned checking that the
random mating assumption often used in mathematical models in population
genetics holds. This implies that the alleles transmitted by each parent are
independent, that is, when there are two possible alleles,
$A$ and $a$ at a locus, with frequencies $p(A)= p$ and $p(a)= 1-p = q$ in the
population, the frequencies of the three genotypes ($AA$, $Aa$ and $aa$)
in the next generation equal $p^2$, $2pq$ and $q^2$. Hardy (\citeyear{1908Hardy}) and
Weinberg (\citeyear{1908Weinberg}) showed that in a large randomly mating population these genotype
frequencies remain the same from one generation to the next. To test whether the Hardy--Weinberg (HWE)
equilibrium holds at a locus, one estimates the frequencies $p$ and $q$ from
a sample of $n$ individuals, using $\bar{p}= [2n(AA)+n(Aa)]/2n$,
$\bar{q}=1-\bar{p}$. Under HWE, the expected genotype frequencies at a particular locus are
obtained by substituting these estimates into the equilibrium distribution.
Then the standard $\chi^2$-test (Gillespie, \citeyear{1998Gillespie}, pages 11--15) is conducted.
When HWE does not hold, different genetic theories and settings typically predict either a decrease or
increase in the number of homozygotes.

An analogous equilibrium distribution holds when there are $k$
possible alleles at a locus and the appropriate $\chi^2$-test is
used. In the highly polymorphic (large $k$) situation, which is of
interest in forensic applications (Evett and Weir, \citeyear{1998Evett}), the
accuracy of the $\chi^2$-test in moderate sample sizes is
questionable; while in studies of rare or endangered species, only
small sample sizes are available (Hedrick, \citeyear{2000Hedrick}, page 74). In the
spirit of Fisher's exact test, Levene (\citeyear{1949Levene}) obtained an exact test
for the number ($h$) of homozygotes that conditioned on the number
of alleles of each of $k$ types. The importance of the problem is
reflected by the current literature developing more computer
intensive exact procedures (Huber et al., \citeyear{2006Huber};
Maurer, Melchinger  and Frisch, \citeyear{2007Maurer});
however, Levene's exact test for HWE was the first. The original
article also derived the large sample distribution of the statistic
and considered the effect of misclassification of a small fraction
of heterozygotes as homozygotes. Finally, Levene expressed the
problem of finding the distribution of $h$ in terms of card
matching; similar analogies between exact tests for HWE and card
shuffling problems are still used today (Weir, \citeyear{1996Weir}, page
110).

A few years later, Levene (\citeyear{1953Levene}) developed the first theoretical
model that examined the effects of spatial variation on fitness
(Hedrick, \citeyear{2000Hedrick}, page  161). During the 1920's Fisher and Haldane asked
an important question: How is polymorphism maintained when selection
is operating?  When there are two alleles at a locus, natural
selection should favor the allele ($A$) most related to survival and
mating, so eventually all the entire population should become
homozygotes ($AA$). As described by Pollak (\citeyear{2006Pollak}), they demonstrated
that each of the two alleles can have a substantial equilibrium frequency
 when heterozygotes are superior in viability to either homozygote and
that a deleterious allele, d, can be maintained at a low equilibrium
frequency due to recurrent mutation of the favored allele to d.
Levene (\citeyear{1953Levene}) showed that two alleles could be maintained when a
population inhabits $K$ ecological niches, migrates between
them, and selection varies among the niches, even if the viabilities of
a heterozygote are between those of homozygotes in all K niches. In
particular, a stable polymorphism can occur when the harmonic mean
fitness of both homozygotes is less than that of the heterozygote.
The basic approach taken by Levene (\citeyear{1953Levene}) is still used in modern
texts (Hedrick, \citeyear{2000Hedrick}, page 161), where references to developments
incorporating genotypic-specific habitat selection, that is, individuals
preferentially migrate to niches in which they have higher fitness
(viability), are described. Recent developments are surveyed by
Hedrick (\citeyear{2006Hedrick}) and Star, Stoffels and Spencer (\citeyear{2007Star}) who investigate the levels of
polymorphism in a model incorporating recurrent mutation and
selection.

In 1960 Professor Howard Levene proposed a now classic test for the
equality of the variances of $k$ populations. The practical
importance of Levene's (\citeyear{1960Levene}) article is demonstrated by the fact
that it has been cited over 1000 times in the scientific literature.
The goal of this paper is to discuss the scientific heritage of
Professor Levene's contribution on both statistical methodology and its
use in a wide variety of disciplines. Other procedures for
testing the equality of variances have been surveyed by Boos and
Brownie (\citeyear{2004Boos}).

Levene's (\citeyear{1960Levene}) original article was motivated by the $k$-sample problem. Before
comparing the sample means, one should check that the underlying populations have
a common variance. At the time, procedures that were easy to calculate were desired.
Section~\ref{sec3} describes the proper use of Levene-type tests as a first stage test
 to select either the standard or Welch-modified $k$-sample ANOVA. With modern
 computers and software, nowadays one can use the Welch method in place
 of ANOVA, as it incurs only a small loss in power when the variances are equal.

 Levene's test, however, remains very useful, as many
scientific questions concern the variances of $k$ populations, rather
than their means or location parameters (centers). For example, to choose among several ways
of delivering the same average dose of a drug, the one with least
variability in the measured dose is preferred. When reviewing the
applied literature, it became apparent that many alternative hypotheses were
best described as a monotonic trend in the variances of the $k$
populations; hence, a modification of Levene-type tests for this situation is proposed.
The increased power of a trend test, which is directed at the alternative of interest,
is illustrated by reanalyzing data from two published studies.

Levene-type tests have become very popular and are used in a wide variety
of applications, for example, clinical data (Grissom, \citeyear{2000Grissom}), marine
pollution (Johnson, Rice and Moles, \citeyear{1998Johnson}), species preservation
(Neave et al., \citeyear{2006Neave}), climate change and geology (\mbox{Henriksen},
\citeyear{2003Henriksen}; Khan, Coulibaly  and Dibike, \citeyear{2006Kahn}; Coulson and
Joyce, \citeyear{2006Coulson}), animal science\break (Waldo and Goering, \citeyear{1979Waldo}; Schom and
Kit, \citeyear{1980Schom}), food quality (Francois et al., \citeyear{2006Francois}), spherical
distributions in astronomy (Fisher, \citeyear{1986Fisher}), regional differences of
semen quality (Auger and Jouannet, \citeyear{1997Auger}), business (Chang, Jain and
Locke, \citeyear{1995Chang}; Christie and Koch, \citeyear{1997Christie}; Plourde and Watkins, \citeyear{1998Plourdes}),
auditing (Davis, \citeyear{1996Davis}), studies of awards in civil cases (Saks et
al., \citeyear{1997Saks}; Robbennolt and Studebaker, \citeyear{1999Robbennolt}; Marti and Wissler,
\citeyear{2000Marti}; Greene et al., \citeyear{2001Greene}), the analysis of data in actual legal cases (Tyler v. Unocal,
304 F.3d 379, 5th Cir. 2002), genetics and evolution (Mitchell-Olds
and Rutledge, \citeyear{1986Mitchell}; Giraud and Capy, \citeyear{1996Giraud}), toxicology (Mayhew,
Comer and Stargel, \citeyear{2003Mayhew}), psychology, education and speech\break (Flynn
and Brockner, \citeyear{2003Flynn}; Cattaneo, Postma and Vechi, \citeyear{2006Cattaneo}; O'Neil, Penrod
and Bornstein, \citeyear{2003ONeil}; Tabain, \citeyear{2001Tabain}), sports (Cumming and Hall, \citeyear{2002Cumming})
and even sex research (Hicks and Leitenberg, \citeyear{2001Hicks}; Hays et
al., \citeyear{2001Hays}).

The original tests along with subsequent modifications that improve the robustness
of the test to non-normality of the underlying data, for example,
Brown and Forsythe (\citeyear{1974Brown}), or improve the statistical performance in certain
 circumstances, for example, unequal sample sizes, are described in Section~\ref{sec1}.
 Section~\ref{sec2} discusses Levene-type tests when the alternative is that the variances of the
 $k$-groups follow a monotonic trend. A modification of the statistic along the lines of the
 Cochran--Armitage trend test,
used to analyze dose-response data, is described. The results of a small simulation study
illustrate its increased power. Our results are consistent with the detailed investigations of
Balakrishnan and Ma (\citeyear{1990Balakrishnan}) and Lim and Loh (\citeyear{1996Lim}) and collectively they provide extensive support
for the use of robust Levene-type tests in practice. Section~\ref{sec3} describes the proper use of
Levene-type tests as a first stage test to decide whether to
analyze the data by the standard or Welch-modified $k$-sample ANOVA. While the two-stage method,
using an appropriate size for a Levene-type preliminary test, remains valid, with modern day statistical
software, in most situations one can use the Welch method, as it is only slightly
less powerful than the standard test when the variances are equal. The use of Levene-type tests in the
analysis of data arising in a wide variety of interesting applications is described in the penultimate
 section (Section~\ref{sec4}). The paper concludes with a summary of recommended
methods and a discussion of topics needing further research.

\section{The Original Test and Further Robust
Modifications}\label{sec1}

A basic problem in ANOVA is to determine whether $k$ populations
have a common mean $\mu$. One has $k$ random samples, $x_{i1},
\ldots, x_{in_i}$, of size $n_i$ from each of $k$ populations with
respective means, $\mu_i$, and variances $\sigma_i^2$,
$i=1,\ldots,k$. The standard $F$-test assumes that in each of the
populations the variable studied has a common variance $\sigma^2$
and compares the between group mean square to the within group mean
square ($s_p^2$), that is,
\begin{equation}
\label{1} F=s_p^{-2}\sum_{i=1}^k(\overline{x}_{i\cdot}-\overline{x}_{\cdot\cdot})^2/(k-1),
\end{equation}
 where $s_p^2$ is the pooled variance, $\bar{x}_{i.}$ is the mean of the
 $i$th group, $\bar{x}_{\cdot\cdot}$ is the grand mean and $N=\sum_{i=1}^k n_i$. It has long
been known that the actual size of the test based on $F$ may differ
noticeably from the nominal size, for example, 0.05, when the groups have
different variances (Sheffe, \citeyear{1959Scheffe}, pages  351--358). This problem is
quite serious when the variances are negatively correlated with the
sample sizes (Krutchkoff, \citeyear{1988Krutchkoff}; Weerhandi, \citeyear{1995Weerhandi}). Hence, it is
important to develop methods for checking the validity of the equal
variance assumption.

Bartlett (\citeyear{1937Bartlett}) proposed a statistic, $M$, for testing the equality of $k$
population variances that is a function of the variances ($s_i^2$)
of the $i$th group. Subsequently, Box (\citeyear{1953Box}) showed that the sampling
distribution of Bartlett's $M$ is not robust to violations of the
assumed normality of the underlying distributions. Box noted that Bartlett's
procedure is more useful as a test of normality than as a test for
equality of $k$ group variances. Box and Anderson (\citeyear{1955Box}) showed that
the effect of normality depends on the kurtosis,
$\gamma_2=\mu_4/\mu_2^2$, the ratio of the fourth central moment of
the underlying distribution to the square of the variance. Assuming
the data from the $k$ groups have the same distribution, the natural
estimator of $\gamma_2$ is
\begin{equation}
\label{2}
\hat{\gamma}_2=\frac{N\sum_{i=1}^k\sum_{j=1}^{n_i}(x_{ij}-\bar{x}_{i\cdot})^4}
{[\sum_{i=1}^k\sum_{j=1}^{n_i}(x_{ij}-\bar{x}_{i\cdot})^2]^2}.
\end{equation}

Multiplying Bartlett's $M$ by $2/(\hat{\gamma_2}-1)$ yields a test statistic,
$B_3$, which has an approximate $\chi^2$-distribution with $(k-1)$
degrees of freedom. Notice that for normal data the expected value
of the factor $2/(\hat{\gamma_2}-1)$ equals 1.0 and as the kurtosis
increases above 3, it becomes {\it smaller}. The statistic $B_3$ is
the form of the Box--Anderson test discussed by Miller (\citeyear{1986Miller});
see also Shorack (\citeyear{1969Shorack}).

In the small samples often encountered in applications of ANOVA, the
higher moments are quite variable, so a test that does not rely on
the fourth sample moment is desirable. To appreciate the idea
underlying the approach adopted by Levene, assume that the group
means $\mu_i$ are known. To measure variance or spread, he
considered various functions of $x_{ij}-\mu_i$, for example,
$|x_{ij}-\mu_i|$ and $(x_{ij}-\mu_i)^2$. The expected value of
$(x_{ij}-\mu_i)^2$  is $\sigma_i^2$, the variance of the $i$th
group, while the expected value of $|x_{ij}-\mu_i|$ is the mean
deviation from the mean, a well-known measure of spread related to a
classical measure of income inequality due to Pietra (Gastwirth,
\citeyear{1972Gastwirth}). Thus, if one knew the group means, one could apply the
standard ANOVA statistic to $|x_{ij}-\mu_i|$ or $(x_{ij}-\mu_i)^2$.

Since the group means, $\mu_i$, are typically unknown, Levene
naturally used the sample group means, $\bar{x}_{i\cdot}$, in their places.
Then $|x_{ij}-\bar{x}_{i\cdot}|$ or $(x_{ij}-\bar{x}_{i\cdot})^2$ are treated as
independent, identically distributed, normal variables, and the usual ANOVA
statistic is utilized. While neither $|x_{ij}-\bar{x}_{i\cdot}|$ nor
$(x_{ij}-\bar{x}_{i\cdot})^2$ is normally distributed, Levene's approach takes
advantage of the fact that classical ANOVA procedures for comparing
means are robust to violations of the assumption that the data
follow a normal distribution (Miller, \citeyear{1968Miller}, page  80). Of course, Levene realized that
$|x_{ij}-\bar{x}_{i\cdot}|$ and $(x_{ij}-\bar{x}_{i\cdot})^2$ are not independent within
each group, as they are deviations from the group mean. However, he
showed that the correlation is of the order $1/n_i^2$ and had the
intuition that this small degree of dependence would not seriously
effect the distribution of the $F$-statistic. After trying different
functions of $(x_{ij}-\bar{x}_{i\cdot})$, for example, square, log etc., Levene
proposed the final version of the test in the form of the classic
ANOVA method applied to the absolute differences between each
observation and the mean of its group  $d_{ij}=|x_{ij}-\bar{x}_{i\cdot}|$,
$i=1,\ldots, k$, $j=1,\ldots, n_i$. Since the $d_{ij}$ are not
normally distributed even when the original $x_{ij}$ are, the
resulting $F$-statistic,
\begin{equation}
\label{3}
F=\frac{N-k}{k-1}\frac{\sum_{i=1}^k(\bar{d}_{i\cdot}-\bar{d}_{\cdot\cdot})^2}{\sum_{i=1}^k\sum_{j=1}^{n_i}n_i
(d_{ij}-\bar{d}_{i\cdot})^2},
\end{equation}
is not exactly distributed as the usual $F$-statistic with $k-1$ and
$N-k$ degrees of freedom. Levene (\citeyear{1960Levene}) showed by simulation that
the usual $F$ statistic provides a good approximation, especially at
the cut-off values corresponding to the commonly used significance
levels, $\alpha=0.01$ and 0.05.

A natural way to increase the robustness of Levene's original
statistic is to replace the group means in the definition of
$d_{ij}$ by a more robust estimator of location, for example, the median
(Brown and Forsythe, \citeyear{1974Brown}) (BFL test). Studies by Conover,
Johnson and Johnson (\citeyear{1981Conover}) and Lim
and Loh (\citeyear{1996Lim}) confirm that utilizing the absolute deviations of the
observations from their group medians, rather than means, is preferable.
Thus, the modern version of Levene's test uses the
$z_{ij}=|x_{ij}-\hat{\mu}_i|$ in place of $d_{ij}$ in~(\ref{3}),
where $\hat{\mu}_i$ are robust estimators of $\mu_i$.

In small samples, for example, when there are no more than 10 observations in each group,
the level of the Levene test can be quite conservative when the group centers are estimated
by their medians. The problem arises from the fact that for odd group sizes, one of the
absolute deviations from the group median must equal 0; and for even sample sizes, two of the
absolute deviations are equal as the group median is estimated by the average of the middle
two observations. Thus, a bootstrap version was proposed by Boos and Brownie (\citeyear{1989Boos}) and shown to have improved power
by Lim and Loh (\citeyear{1996Lim}). An alternative modification was suggested by  Hines and
Hines (\citeyear{2000Hines}). When the number of observations $n_i$ in the $i$th group
is odd, they propose to remove a structural zero $z_{im}$ for
$m=[n_i/2]+1$ (here $[y]$ is the floor function of $y$); when $n_i$ is
even, then the two smallest and necessarily equal deviations
$z_{i[n_i/2]}$ and $z_{i[n_i/2+1]}$ are replaced by one single value
$\sqrt{2}z_{i[n_i/2]}$. The Hines--Hines (\citeyear{2000Hines}) procedure increases the
variability of $z_{ij}$, reducing degrees of freedom by one for each group to
compensate for  the structural zeros as well as decreasing the Error Sum of
Squares and Mean Squares in the Levene ANOVA table. As a result, this simple
modification provides a test with size closer to the nominal one, especially
in small samples. In addition, this usually provides a Levene-type test with increased power.

Several authors, Martin and Games (\citeyear{1977Martin}),\break O'Brien (\citeyear{1979OBrien}), Keyes and
Levy (\citeyear{1997Keyes}) and O'Neil and Mathews (\citeyear{2000ONeil}, \citeyear{2002ONeil}), examined the effect
that unequal sample sizes create when the data follows a normal
distribution and proposed appropriate correction factors. In the
one-way ANOVA, under $H_a$, the variances of the observations
$\sigma_i^2$ differ, implying that the expected values of the
$d_{ij}$ are given by
\begin{equation}
\label{4}
E(d_{ij})=\sigma_i\sqrt{\frac{2}{\pi}\biggl(1-\frac{1}{n_i}\biggr)}.
\end{equation}
Notice that equation~(\ref{4}) implies that even under $H_0$, that is,
when all groups have a common variance $\sigma^2$, the expected
group averages differ. Thus, large differences in the sample
sizes, $n_i$, may cause the original Levene test to reject the null
hypothesis when it is true.

O'Brien (\citeyear{1979OBrien}) and Keyes and Levy (\citeyear{1997Keyes}) remove this design effect
by replacing $d_{ij}$ by $u_{ij}= d_{ij} /\sqrt{1-1/n_i}$, which
have the same expected value and are proportional to the absolute
values of the standardized residuals from the original ANOVA. Then
one applies OLS ANOVA to the $u_{ij}$.  O'Neill and Mathews (\citeyear{2000ONeil})
obtained the covariance matrix of $u_{ij}$ and created the
appropriate weighted least squares estimates of the within group
and between group variances of $u_{ij}$ and obtained the
corresponding $F$-test. When the $n_i$ are equal, to $n$, they showed that the
weighted $F$-statistic is a factor, $m$, times the OLS $F$-test. Furthermore,
$m$ tends to 1 as $n$ increases. O'Neill and Mathews (\citeyear{2000ONeil}) also
obtained the corresponding multiplier when deviations from the group
medians are used. Manly and Francis (\citeyear{2002Manly}) showed that
when the significance level of the $F$-test was determined by
randomization of the residuals of deviations from the sample
medians, it was very robust to nonnormality and was less affected
by modest differences in the $n_i$.

\section{Levene-Type Tests for a Trend in the Group Variances}\label{sec2}

While reviewing the large number of studies applying Levene's test
or the Brown--Forsythe modification, we noticed that the alternative hypothesis appropriate
to the subject matter often indicated that the variances would follow a decreasing or
increasing trend; for example, the groups might correspond to dose levels or
could be classified by status on a monotonic scale. It is well known
that tests directed at a specific alternative typically are more
powerful in detecting a particular alternative (Agresti, \citeyear{2002Agresti}; Freidlin and Gastwirth, \citeyear{2004Freidlin}).
Often, under the alternative the $k$ groups can be arranged so that their variances
increase, that is, $H_a$ is  $\sigma_1<\sigma_2<\cdots<\sigma_k$.
A number of procedures which employ the idea of regressing
the sample variances of each group vs. some preselected scores or considering a particular
contrast have been developed for this problem (Vincent, \citeyear{1961Vincent}; Chacko, \citeyear{1963Chacko};
Fujino, \citeyear{1979Fujino}; and Hines and Hines, \citeyear{2000Hines}). Here we follow
the simple linear regression approach in which
scores $w_1<w_2< \cdots <w_k$ are assigned to each observation in
the $i$th group ($i=1,\ldots,k$). The expected value of the slope $\hat\beta$~(\ref{5eq})
of the regression line relating the $z_{ij}$  to the $w_i$ is zero under the null hypothesis,
  but will be positive (negative) under the alternative that there is an increasing (decreasing)
  trend in the variances. The estimator $\hat{\beta}$  of $\beta$ is given by
\begin{eqnarray}
\label{5eq} \hat{\beta} & = & \frac{\sum_{i=1}^k
n_i(w_i-\bar{w})(\bar{z}_{i\cdot}-\bar{z}_{\cdot\cdot})}{\sum_{i=1}^k n_i(w_i-\bar{w})^2},
\\
 \bar{w}&=&\sum_{i=1}^k n_iw_i/N, \nonumber
\end{eqnarray}
where $\bar{z}_{i\cdot}, i=1,\ldots, k,$ are the group means of $z_{ij}$ and $\bar{z}_{\cdot\cdot}$ is the grand
mean over $\bar{z}_{i\cdot}, i=1,\ldots, k$. When the
observations in each group come from a normal
distribution, the null hypothesis that the group variances are equal
implies that the mean deviations from the group means (or medians)
also are equal. When the variances or other measure of spread are
equal, $\hat{\beta}$ should be centered around zero, while under the
alternative that the group variances increase $\hat{\beta}$ should
be positive.

The expression for the slope $\hat{\beta}$ in~(\ref{5eq}) is analogous
to the classic one degree of freedom test for the strength of linearity (Johnson and
Leone, \citeyear{1964Johnson}, page 78) or the Cochran--Armitage trend test for binary
data (Piegorsch and Bailar, \citeyear{2005Piegorsch}) and its numerator is like a
covariance between the group centers $\bar z_{i.}$ and scores $w_i$.
Hines and Hines (\citeyear{2000Hines}) show that using contrasts that reflect the
alternative or suspected trend have higher power than
the usual $F$-statistic~(\ref{1}) for homogeneity applied to the
$z_{ij}$. Abelson and Tukey (\citeyear{1963Abelson}) showed the linear scores are
efficiency robust over a wide range of increasing trends, so they are
commonly used. If the alternative hypothesis implies a specific
nonlinear trend, one should use the corresponding values for $w_i$, for example, $w_i=i^2$
or $w_i=\sqrt{i}$. Roth (\citeyear{1983Roth}) and Neuhauser and Hothorn (\citeyear{2000Neuhauser}) developed trend tests
using order-restricted inference. These methods may be more powerful when the trend is
monotonic but far from linear, they are not explored here. The increased power
 of Levene-type trend tests will be seen in Section~\ref{sec4} where we reanalyze data
sets from two scientific studies.

\begin{Remark*} If the true group centers are known, then the
standardized Levene-type trend statistic asymptotically follows
a standard normal distribution, as do results from Proposition 2.2 of
Huber (\citeyear{1973Huber}), Theorem 1 of Arnold (\citeyear{1980Arnold}) and Carroll and Schneider (\citeyear{1985Carroll}).
In practice, however, the ``true'' group centers are
typically unknown and estimated from a sample of observations.
In the one-sample setting Miller (\citeyear{1968Miller}) showed that Levene's
original statistic, using absolute deviations from the group means,
is asymptotically distribution-free only when the underlying
distribution is symmetric; if the sample group median are employed, then
the statistic is asymptotically distribution-free. The corresponding large sample result for $k$
groups was proved by Carroll and Schneider (\citeyear{1985Carroll}). Using the results
of Carroll and Schneider (\citeyear{1985Carroll}), Bickel (\citeyear{1975Bickel}) and Carroll and
Ruppert (\citeyear{1982Carroll}), it can be shown that if the ``true'' group centers are
unknown, then the size of Levene's trend statistic 
determined from its asymptotic distribution is correct only when the
group location parameters are estimated by the group medians.
\end{Remark*}

A small simulation study considering samples from
normal and heavy-tailed symmetric distributions was conducted where a robust trimmed
mean (Crow and Siddiqui, \citeyear{1967Crow}; Gastwirth and Rubin, \citeyear{1969Gastwirth}; Andrews et
al., \citeyear{1972Andrews}), the average of the middle 50\% of the data, was also used
to estimate the group centers. Our simulation study\footnote{All
calculations are performed using the R package \textit{Lawstat} that is
freely available from
\href{http://cran.r-project.org/}{http://cran.r-project.org/}.} indicates that for small
and moderate sample sizes, the 25\% trimmed versions of Levene's ($L_{0.25}$)
trend tests yield the most accurate size for a test at the nominal 5\% level for all
the distributions (normal, exponential, $t$- and $\chi^2$-distributions with 3 degrees
of freedom) studied. In contrast, the corresponding test statistics using the sample
means have levels exceeding the nominal 5\%, especially for the heavy tailed and skewed
distributions. Using medians, as in the Brown--Forsythe version, substantially
underestimates the size of the test for small samples, especially for normal data.
Overall, all the three versions of Levene's trend test, that is, the mean, median and
25\% trimmed mean based, were more powerful against monotonic trend alternatives than the
corresponding homogeneity tests, especially for small sample sizes. This is true even when
the scores differ somewhat from the true trend, for example, the
  linear scores 1, 2, 3 are used when the ratios of the standard deviations are $1\dvtx 3\dvtx 5$.
  As expected, in larger samples the difference in performance between Levene-type
  homogeneity and trend tests is minor.

\begin{table*}[b]
\tablewidth=301pt
\caption{The actual sizes of a nominal 0.05 level test
for the three procedures. The results are based on 10,000
simulations}\label{T7}
\begin{tabular*}{301pt}{@{\extracolsep{4in minus 4in}}lcccccc}
\hline
 & \multicolumn{3}{c}{\textbf{10, 10, 10}} & \multicolumn{3}{c@{}}{\textbf{10, 10, 20}}
\\[-6pt]
& \multicolumn{3}{c}{\hrulefill} & \multicolumn{3}{c}{\hrulefill}
\\
$\bolds{\sigma_1: \sigma_2:\sigma_3}$ & $\bolds{1:1:1}$ & $\bolds{1:2:3}$ & $\bolds{1:3:5}$ &
$\bolds{1:1:1}$ &  $\bolds{1:2:3}$ &
\multicolumn{1}{c}{$\bolds{1:3:5}$}
\\
\hline
ANOVA  & 0.0481 &  0.0665 &  0.0665 &
0.0512 &  0.0264 &  0.023\phantom{0}
\\
 Welch ANOVA & 0.0485 &  0.0518&   0.053\phantom{0} & 0.0514&
0.0524 &  0.0529
\\
 Adaptive ANOVA &  0.0496&   0.0572 &  0.0539 & 0.0546& 0.0514
&  0.0529
\\[5pt]
 & \multicolumn{3}{c}{\textbf{10, 20, 10}} &
\multicolumn{3}{c@{}}{\textbf{20, 10, 10}}
\\[-6pt]
& \multicolumn{3}{c}{\hrulefill} & \multicolumn{3}{c}{\hrulefill}
\\
 $\bolds{\sigma_1: \sigma_2:\sigma_3}$ & $\bolds{1:1:1}$ & $\bolds{1:2:3}$ & $\bolds{1:3:5}$ &
$\bolds{1:1:1}$ &  $\bolds{1:2:3}$ &
\multicolumn{1}{c}{$\bolds{1:3:5}$}
\\
\hline
ANOVA  & 0.0491 & 0.0714  &0.0867&
0.0542 & 0.1212 & 0.1399
\\
 Welch ANOVA & 0.0494 & 0.0495 & 0.0524  & 0.0557&
0.0506 & 0.0515
\\
 Adaptive ANOVA & 0.0523 & 0.0554 & 0.0528 & 0.0572&
0.0564 & 0.052\phantom{0}
\\
\hline
\end{tabular*}
\end{table*}

\section{Using Levene's Test as the First Stage in
Adaptive ANOVA Tests}\label{sec3}

In many applications adaptive procedures that utilize
a preliminary test to choose the estimator or test for the final
analysis improve the accuracy of the final inference (Hall and
Padmanabhan, \citeyear{1997Hall}; O'Gorman, \citeyear{1997OGorman}). For example, Hogg (\citeyear{1974Hogg}) and
Hogg, Randles and Fisher (\citeyear{1975Hogg}) use a measure of tail-weight to
select the estimator of the location parameter; Freidlin, Miao
and Gastwirth (\citeyear{2003Freidlin}) use the $p$-value of the Shapiro--Wilk test to select a powerful
nonparametric test for the analysis of paired differences. Miao and Gastwirth (\citeyear{2009Miao}) use
the ratio of two~measures~of spread~to~choose the
nonparametric test to analyze paired data for the second stage. These
methods have been successful in the one-sample problem because
heavy-tails can severely affect the behavior of the sample mean and
an appropriate preliminary test enables one to choose a robust
estimator or test that has high efficiency across a class of
distributions with tail weight close to that of the sample.
Recently, Schucany and Ng (\citeyear{2006Schucany}) noted that preliminary tests must
be used with care, as at the second stage,  the analysis is
{\it conditional} on the results of the first-stage test. They
demonstrated that graphical diagnostics for normality are preferable
to a formal test of normality at the first stage when the objective is
to make inferences about the population mean.

For testing the equality of $k$ sample means, when the variances may not be equal,
Welch (\citeyear{1951Welch}) provided the
following modification of the usual ANOVA $F$-test:
%
\begin{eqnarray}
\label{6}
 F_W&=&\biggl(\sum_{i}w_i(\bar{x}_{i\cdot}-\hat{x})^2/(k-1)\biggr)\nonumber
\\[-2pt]
&&{}\bigg/\biggl[1+\frac{2(k-2)}{k^2-1}
\\[-2pt]
&&{}\hspace{32pt}\times \sum_{i}\frac{1}{n_i-1}
 \biggl(1-\frac{w_i}{\sum_j w_j}\biggr)^2\biggr],\nonumber
\end{eqnarray}
where $w_i=n_i/s_i^2$ and $\hat{x}=\sum w_ix_i/\sum w_i$.

This Welch modification rejects the null hypothesis of equal means
if the $F$ statistic (\ref{6}) is larger than the critical value determined from an $F$
distribution with degrees of freedom $f_1^*$ and $f_2^*$, where
\begin{eqnarray}
\label{7}
 f_1^*&=&k-1,\nonumber
\\[-9pt]\\[-9pt]
f_2^*&=&\biggl[\frac{3}{k^2-1}\sum_i\frac{1}{f_i}
\biggl(1-\frac{w_i}{\sum_jw_j} \biggr)^2\biggr]^{-1}. \nonumber
\end{eqnarray}

When $k$ is 2, the procedure reduces to the Welch \citeyear{1938Welch} two-sample
$t$-test. Because the test using (\ref{6}) allows for \mbox{unequal}
variances, one needs to examine whether it incurs a noticeable
loss of power when the group variances are equal. This section reports the
results of a small simulation study that compares three tests: the
usual ANOVA $F$-test, the Welch modification (\ref{6}) and an adaptive
ANOVA. The adaptive procedure is the following: first use a
Levene-type test to see whether the variances are equal or not. If
the test concludes that the variances are equal, use the ordinary
ANOVA $F$-test, otherwise, use the Welch modification. The results indicate that just using the Welch
method (\ref{6}), which is now available on statistical packages, is easier to use than the adaptive ANOVA
and only incurs a small loss in power when the variances are equal.

The study focused on testing whether the means from three
\textit{normal} distributions are equal. Following the
recommendations of Bancroft (\citeyear{1964Bancroft}) and Huber (\citeyear{1972Huber}) that the level
of a preliminary test should be greater than 5\%, a level of 15\% is
used here.

Table~\ref{T7} shows the observed level of the three tests for different sample sizes and different variance ratios.
The nominal level is 5\%. Clearly, the Welch adjusted ANOVA test and the adaptive procedure
preserve the nominal levels very well for all sample sizes and
variance ratios studied. These results are consistent with previous studies of the two-sample
situation (Moser, Stevens and Matts, \citeyear{1989Moser}, \citeyear{1992Moser}; Weerhandi,
\citeyear{1995Weerhandi}; Zimmerman,
\citeyear{2004Zimmerman} and Vangel, \citeyear{2005Vangel}). In contrast, the actual level of the ordinary ANOVA $F$
test is affected when the variances are not equal. In some situations,
the actual size of the test can be as large as 0.1399, for example, when $(n_1,n_2, n_3) =(20,10,10)$ and
$(\sigma_1\dvtx
\sigma_2:\sigma_3)=(1\dvtx 3\dvtx 5)$.

The powers of the adaptive and Welch ANOVA tests were also investigated by simulation.
When the variances are equal, the powers of the
adaptive procedure are about 2--3\% higher than the Welch adjusted
ANOVA $F$-test. When the variances are not equal, the Welch adjusted
test has higher power, about 2--3\% more than the adaptive one.
Overall, the difference in power between the two
procedures is quite small, rarely more than 0.02. (Detailed results
can be obtained from the authors.) Thus, both the Welch method and the adaptive ANOVA
are valid procedures.

The results reported in Table~\ref{T7} use the group medians to estimate their centers,
in the preliminary Levene-type test. Simulation studies, using the
25\% trimmed means in place of the medians in the Levene test, yielded similar
results. Other simulations explored the role of the size of the
preliminary test. The findings indicate that the size of the
first-stage test should be in the range 15\% to 25\% in order for
the adaptive procedure to have the nominal size (0.05) and have
reasonable power. These results confirm the recommended levels of
25\% by Bancroft (\citeyear{1964Bancroft}) or 20\% by Huber (\citeyear{1972Huber}, \citeyear{1973Huber}) for the size
of a preliminary test.

Both the Welch and the adaptive tests are more robust to departures from
the equal variance assumption than the usual ANOVA $F$-test. These
two tests are nearly as powerful as the standard $F$ test when the
group variances are equal. As the Welch test is simpler, we recommend it for general use.
 Researchers in areas where the two-stage method is commonly accepted, however,
 can still rely on it. The size of the Levene-type preliminary test should be
 between 15\% and 25\%.

\section{The Wide Applicability of Levene's Test and Its Modifications}\label{sec4}

The important role statistical design, methodology and inference
have in a wide array of intellectual disciplines is exemplified by
the numerous applications of Levene-type tests. This section
describes how Levene-type tests were used in a number of interesting
studies from a variety of disciplines. In many cases the Levene-type test
was used as a preliminary check of the equal variance assumption in classical ANOVA; in
others, the scientific issue concerned the equality of the variances of measurements from
$k$ populations. The topics described were chosen from hundreds of valuable scientific
contributions and illustrate the broad scientific impact of Professor Levene's method.

\subsection{Applications in Archeology and Ethnography}

Archaeologists are concerned with the effects increasing
economic activity has on older civilizations. Economic growth encourages
specialization in the production of goods, which led to the ``standardization
hypothesis,'' that is, increased production of an item would lead to its
becoming more uniform.  Kvamme, Stark and Longacre (\citeyear{1996Kvamme}) tested this theory on a type of
earthenware, chupa-pots, from three Philippine communities that
differ in the way they organize ceramic production. In Dangtalan,
pottery is primarily made for household use and restricted exchange.
Dalupa has an extensive nonmarket based barter economy, where part-time
specialist potters trade their output for other goods. The
village of Paradijon is near the Provincial capital; full-time pottery
specialists sell their output to shopkeepers, located in the village or in the
capital, for sale to the general public. To test the
``standardization''
hypothesis, these authors took measurements on three characteristics (aperture, circumference and
height) of two-chalupa pots from the three areas and used the $F$-test
and Brown--Forsythe version of Levene's test to compare the variation
among pots produced in each area. The null hypothesis is that the
variance or spread of each characteristic is the same in the three
areas, while the alternative is that they differ.

After demonstrating that typically the measurements did not follow a
normal distribution and had heavier tails, the authors showed
(their Table 5) that the usual $F$-test can yield substantially
different $p$-values than those obtained from Levene's test. For example,
comparing the {\it circumference} of the 55 pots from
\mbox{Dangtalan} with 170 from Dalupa, the standard $F$-test statistic yielded
1.24, leading to acceptance of the null hypothesis that variances are the same. In
contrast, the robust Levene test yields a $p$-value $=$ 0.001.
Several other pair-wise comparisons showed that the $F$-test could yield much lower
$p$-values than the robust Levene method. Here we apply the three
Levene type tests for homogeneity of variances described in Section
\ref{sec2} to assess whether the variances of the \textit{apertures} of the
two-chalupa pots from the three locations are the same. All three tests,
the original Levene's test (L), the Brown and Forsythe version (BFL) and the trimmed version ($\rm{L}_{0.25}$), conclude that the
variation in each of the three measured characteristics of the pots made in the regions
are statistically significant. These results provide support for the standardization hypothesis.

The standardization hypothesis predicts that as eco\-nomies develop,
production intensifies, causing products to become more uniform or
less variable. A test having high power for this particular
alternative hypothesis, that is, the standard deviation of the three
characteristics of the pots should \textit{decrease} with {\it
increasing} economic development, is preferable to a general test of homogeneity of
the variances. Because the alternative hypothesis predicts that the variances of the
three characteristics in pots from Dangtalan should be larger than those produced in Dalupa, which
in turn should be larger than pots made in Paradijon, we analyze the data with the
trend test~(\ref{5eq}).

To appreciate the increased power of the directed trend test, we
analyzed the aperture data, kindly provided by Professor Kvamme. Using
weights 1, 2 and 3 and deviations from the group means, mid-means and medians,
respectively, in~(\ref{5eq}) yielded $p$-values 0.0001,
0.0004 and 0.0004 respectively. The estimates of the slope
$\hat{\beta}$ were similar: $-$1.77, $-$1.68 and $-$1.81. All
three $p$-values are less than one-half those obtained
from the corresponding test of homogeneity and provide stronger evidence
in favor of the ``standardization hypothesis.''

\subsection{Applications in Environmental Sciences}

Even before Katrina, ecologists studied the effect of hurricanes on
forests, especially their rejuvenation after a severe storm. The
catastrophic uprooting of trees creates mounds, pits and other
micro-sites that provide possible locations for a particular species
to regenerate. Carlton and Bazzaz (\citeyear{1998Carlton}) simulated the effect of a
hurricane by pulling down selected canopy trees and then measuring
several important environmental resources (soil organic matter
concentration, nitrogen transformation rates and the amount of $\mathrm{CO}_2$)
at five types of micro-sites that are created after a storm. These are as follows: mounds;
pits; top sites, which are north facing forest floor surfaces; open sites, which are
level and unshaded portions of the forest floor; and level portions of the forest floor
that are covered by ferns or similar vegetation, called fern sites. For comparative
purposes, measurements of the various resources were taken in a
control area. Several questions were addressed, including: what were the residual
effects of the disturbance on the average levels of key resources
in the disturbed sites three years later? Did the simulated hurricane
increase resource heterogeneity among the different micro-sites?

One-way ANOVA was used to test the differences in the average level of a resource among the five
types of micro-sites. Samples of size five were taken from eight different micro-sites of each
type. The authors applied the original version of Levene's test
to check whether the variances of the measurements in the five groups were equal.
When it indicated unequal variances, a single degree of freedom
contrasts (SDFC) were used in lieu of ANOVA (Milliken and Johnson,
\citeyear{1984Milliken}). When the homogeneity of variances assumption was satisfied
and the ANOVA indicated significantly different effects among the
micro-sites, a standard multiple comparison method for contrasts was
utilized.

Due to nonhomogeneity of variance, Carlton and Bazzaz (\citeyear{1998Carlton}) needed
to use an SDFC to establish that the top sites were higher in soil
organic matter than all other micro-sites, while percent
soil water by mass was highest on fern, open and control sites.
 The standard ANOVA method was applicable to the data on
climate factors. The $\mathrm{CO}_2$ concentration was lowest on mounds.
 A major finding was that photon flux density (PFD), a
measure of the amount of light level, on mounds, open sites and pits
was \textit{higher} than in the control (undisturbed) area. In
contrast, the PFD on fern and top micro-sites was less than in the
control area. The results suggest that hurricanes increase light levels
immediately, which may encourage the growth of shade-intolerant species,
while the change in the availability of various soil resources is more gradual. The authors
carefully noted that their simulation cannot replicate all the
features, for example, very high winds, of a real hurricane. Presumably,
similar studies are underway in the areas most affected by the
recent severe storms to assist in the regeneration of plant species.

\subsection{Applications in Business and Economics}

The problem of comparing $k$ sample variances
also arises in business and economics. Here, two
applications of Levene's test in this area are briefly described, although there
are many other interesting studies (Davis, \citeyear{1996Davis};
Christie and Koch, \citeyear{1997Christie};\break
Dhillon, Lasser and Watanbe, \citeyear{1997Dhillon};
Chang, Pinegar and Schacter, \citeyear{1997Chang};
Koissi, Shapiro and Hognas, \citeyear{2006Koissi}) that implemented the
procedure.

Prior to the 1970s, the price of oil was less variable than that of
other commodities; first due to the dominance of the major oil
companies and later the formation of OPEC by the main countries
producing it. To examine whether the behavior of oil prices changed
in the 1980s and became more similar to that of other commodities,
which tend to have large price fluctuations, Plourde and Watkins
(\citeyear{1998Plourdes}) applied Levene's test to monthly price changes, measured by
the logarithm of the ratio of the price in the current month to that of
the previous month, in oil and other commodities (tin, zinc, wheat, etc.). After noticing that the
monthly price changes of the two oil markets (West Texas and Brent)
and the seven other commodities have high kurtosis, the authors
realized that the usual assumption that the underlying populations
all have the same shape or distribution and differ only in the scale
parameter was implausible. Thus, they used both the Brown--Forsythe adaptation of
Levene's test and the nonparametric Fligner--Killeen (\citeyear{1976Fligner}) test in a
series of pairwise comparisons to assess the relative dispersions of the price
changes. In general, both tests showed that the monthly oil price changes
were statistically significantly more dispersed than those of other
commodities, except for lead and nickel, during the years 1985--1994.
The modified Levene test did detect an increase in the dispersion of the price
changes of zinc that the F--K test did not. This is consistent with the findings of
Algina, Olejnik and Ocanto (\citeyear{1989Algina}), indicating that the
O'Brien (\citeyear{1979OBrien}) and BFL tests have relatively high power
and preserve the nominal significance for the family of distributions and sample sizes
they studied.

Stock market analysts and investors are interested in deciding
whether various actions by companies assist them in predicting the
future earnings and market prospects of those firms. Sant and Cowan
(\citeyear{1994Sant}) studied the impact of an omission of a dividend by a company
on the variability of both the forecasts of future earnings and the
actual earnings. They compared the earnings and forecasts of
companies that omitted a dividend during the period 1963--1984 by
comparing the variances of the actual or forecasted earnings per
share two years after the omission and two years before. Since the
data was not normal, they utilized a robust Levene test (BFL). All
comparisons showed that the variability of actual and forecasted
earnings were significantly larger after the dividend omission. The
authors also were careful to construct a control group of similar
firms that did not omit a dividend. In a similar comparison, the
earnings of these companies was not significantly greater in the later
period. Because the increased earnings variability only
occurred in the firms that omitted a dividend, their findings support
the hypothesis that managers omit dividends when a firm's earnings
become less predictable.

\subsection{Applications in Medical Research}

Since a cancer patient's probability of survival is increased when the
disease is detected at an early stage, screening tests are an
essential part of health care. Women over 50 typically have a
mammogram every year or two.  In many European nations, for example, the
UK, mammograms tend to be evaluated at a few central locations, so
each radiologist reviews many of them. In contrast,
the system in the US is more decentralized, so there are fewer
radiologists who assess a large number of mammograms. To study
whether the accuracy of the mammogram is related to the volume a
radiologist sees, Esserman et al. (\citeyear{2002Esserman}) obtained a sample of 59
radiologists in the US and 194 high-volume radiologists in the
UK The number of US radiologists in each volume category was 19
low ($<$100 per month), 22 medium (101--300) and 18 high ($>$300).
Each radiologist was given a test set of 60 two-view films that
contained 13 cancers.

In the disease screening context (Gastwirth, \citeyear{1987Gastwirth}; Pepe, \citeyear{2003Pepe})
accuracy is measured by both sensitivity (the probability a person
with cancer is correctly identified) and specificity (the
probability a healthy person is correctly classified). One can
increase the sensitivity of a screening test by lowering the
threshold level for classifying a subject as diseased, which
decreases the corresponding specificity. A radiologist's accuracy is
evaluated by their sensitivity at a specificity level of 0.90.
Therefore, the authors fit an ROC curve (Gastwirth, \citeyear{2001Gastwirth}; Pepe,
\citeyear{2003Pepe}) to the data for each radiologist using a variant of the
binormal model (Dorfman and Berbaum, \citeyear{2000Dorfman}). For the
US radiologists, average sensitivity was 70.3\% for
those in the low-volume category, 69.7\% for the medium volume group and 77\% for
readers of a high-volume of mammograms. High-volume UK radiologists had an average
sensitivity of 79.3\%. Because the BFL test indicated that the variances in
the sensitivities of the radiologists in the groups were not equal, separate pairwise
Welch-type $t$-tests were performed and showed that the differences among
the average sensitivities were statistically significant. The area under
the ROC curve (AROC) was used as a second measure of accuracy. The areas
under the ROC curve ranged from an average of 0.832 for low-volume
readers to 0.902 (0.891) for high volume UK (US) radiologists. Levene's test
showed that the variances of the AROC in the four groups were
statistically significant. Thus, Bonferroni adjusted pairwise comparisons were carried
out and showed that the high volume radiologists were noticeably
more accurate than the low and medium volume readers. Several related comparisons were conducted,
which confirmed that the percentage of \mbox{cancers} detected by
high volume radiologists significantly exceeded the corresponding
percentage detected by lower volume \mbox{radiologists}. Their finding that higher volume
improves diagnostic performance suggests that the quality and efficiency
of screening programs can be improved by reorganizing them into more centralized
high-volume centers.

Berger et al. (\citeyear{1999Berger}) utilized a database of 6026 echocardiograms
that were read by one of three similarly qualified readers to assess
the differences in frequency of several diagnoses and related
measurements. The numbers of echocardiograms read by the readers (1,
2, 3) were 2702, 2101 and 1223, respectively. Levene's test was used
to assess the variability in the measurements of several continuous
characteristics, of which we discuss two: left atrial dimension (LAD) and
left ventricle ejection fraction (LVEF). The median values of LAD for the
three readers were as follows: 3.9, 3.9 and 3.8, respectively. The Kruskal--Wallis test (K--W test),
 however, showed that the three groups were significantly different, but the Median
test did not detect any difference. Levene's test
indicated statistically significant differences in the variability
of LAD measurements made by the three doctors. Like the Wilcoxon test, the
null distribution of the K--W test is affected by differences in the scale parameters or variances
of the underlying distributions. The investigators may not have been aware of this issue and did
not explore whether the differences among the variances of the three distributions would
be sufficient to change the inference obtained from the usual K--W
test.

 The median values of the LVEF measurements made by the three readers were identical, 57.5 and Levene's
test found no difference in their variability. A somewhat surprising statistically
 significant difference in location was found by both the
Kruskal--Wallis and the Median tests. This might be due to the large,
 but unequal, sample sizes and/or the fact that the LVF
measurements appear to be left-skewed, as the mean values of all
three readers (52.7, 51.5 and 51.6) were less than the corresponding
medians. The nonnormality and skewness of both data sets were
indicated by Q--Q type plots. In contrast to the LVF data, the LAD measurements appear
to be right skewed, with a fairly heavy right-tail.

A major finding was that the prevalence of mitral valve prolapse
(MVP) differed in the three groups (5.3\%, 3.0\% and 4.8\%), as did
the recognition of clots (1.9\% for reader 1 versus about 0.5\% for
readers 2 and 3). After checking that the individuals in the three
groups had similar age and sex compositions, the authors noted that
these differences would be difficult to detect in a typical
small-scale reproducibility study. The data used in this study, as in many epidemiologic investigations,
were observational, and not obtained from a randomized clinical trial. Thus, a sensitivity
analysis based on generalizations of Cornfield's inequality (Rosenbaum, \citeyear{2002Rosenbaum}) can be
used to assess whether an omitted variable could explain the
observed differences in the prevalence of heart problems found by
the three readers. The article noted that some data was missing in a
small proportion of cases but, given the large sample size, the authors
decided not to impute those data. In this particular case, they are
probably correct, however, from a statistical viewpoint it would be
preferable for researchers to report the proportion of missing data.
Then readers could assess whether it might affect the results. For
example, the Kruskal--Wallis test of equality of the location
parameters of the LVF measurements just reached statistical
significance at the 0.05 level. If the proportion of missing
measurements varied among the three readers, then the data would not
be consistent with ``missing at random'' and the significance of the
data might change with the method of imputation adopted.

An interesting study (Rosser, Murdoch and\break Cousens, \citeyear{2004Rosser})
demonstrated that a medical problem, optical defocus, increases the
variability of the measurements of visual acuity. When visual acuity
is repeatedly measured on the same person, the
recorded scores can vary. This test-retest variability (TRV) is measured
in units of the logarithm of the minimum angle of resolution (logMAR) and is a
form of measurement error. Previous studies yielded estimates of the
95\% range of TRV measurements between  $\pm0.07$ to $\pm0.19$
logMAR. Following up on a conjecture that the length of the 95\% TRV
range might increase with the amount of defocus, these investigators
examined 40 subjects under three conditions: no defocus or full
refractive correction, full correction plus 0.50 D and full
correction plus 1.00 D. The order of the six measurements given to a
participant was randomized and no eye chart was used for consecutive
measurements. When the same chart was used the patient was asked to
read it forward one time and backward on the other. Thus, memory or
learning as well as the potential effect of fatigue were controlled
for in the experimental design. Following a common practice in
ophthalmology of ignoring the matching, the authors applied the
original Levene test of homogeneity of variances and obtained a
significant result ($p=0.00023$). The trend test using the group
means yielded a more significant result ($p = 4.16\times 10^{-5}$).
Similarly, the trend test using group medians yielded a lower
$p$-value than the test of homogeneity (0.00024 vs. 0.00124). As expected,
the $p$-values obtained using the 25\%-trimmed means of each group as their centers
were in between those obtained using the mean and median. The smaller $p$-value of the trend test,
which is directed at the alternative of interest, provides greater support
for the conclusion that the variability of measured visual acuity increases
with the degree of optical defocus than the test of homogeneity.

\subsection{Applications in Legal Studies and Law Cases}

In product liability and other tort cases, there is concern that
monetary damages are not proportionate to the actual harm. Furthermore, individuals
who contract the same illness after exposure to the same toxic product can receive
very different monetary compensation from the legal system. Since the
deliberations of actual jurors are confidential, researchers (Saks
et al., \citeyear{1997Saks}; Goodman, Green and Loftus, \citeyear{1989Goodman}; Robbennolt and Studebaker, \citeyear{1999Robbennolt};
Marti and Wissler, \citeyear{2000Marti}) have varied the scenario described or the
instructions given to mock jurors to evaluate whether the
variability of awards for similar injuries can be reduced.

For example, Saks et al. (\citeyear{1997Saks}) explored the effect of giving jurors
different types of information to guide their awards. Thus, some
jurors were given no guidance (control), some the average award for
the type of injury, some a range or interval of values, some both an
interval and the average, and some were given some examples of awards in
similar cases while some were given a cap or upper limit. These
researchers also varied the severity of the injury. For low severity
injuries, Levene's original test yielded a highly significant result
$F_{(5,114)} =11.5$, ($p<0.001$). Significant variation also
occurred in the medium and high injury categories. Somewhat
unexpectedly, jurors given a cap had the most variable awards for
low-level injuries. In the high-level category, the most variable
conditions were the ones when no guidance or just the average award
was provided to the mock jurors. Robbennolt and Studebaker (\citeyear{1999Robbennolt})
explored the effect of varying the cap on punitive damage awards.
Levene's test showed that the variability of those awards also
increased with the size of the cap the mock jurors were given,
however, the variability of the awards the control or no cap mock
juries gave was less than those of mock juries given the highest
cap (\$50 million). These authors also showed that overall
variability of jury awards was reduced when the awards for
compensatory damages and punitive damages were made in two separate
stages of jury deliberation.

The \emph{Tyler v. Union Oil Co. of California} (304 F. 3d 379, 5th Cir.
2002) case concerned age discrimination in layoffs. First,
plaintiffs' expert showed that recent job evaluations received by
employees and their retention status were not significantly
correlated. Then he compared the age distribution of the employees
who were terminated to those who were retained in various locations
of the firm. Levene's test was used to determine whether the usual
$t$-test, which assumes the variances of the distributions are
equal, or the Welch modified $t$-test is more appropriate. In most
comparisons both versions of the $t$-test were significant. In one
location, Ponville, the ages of 36 employees who were placed in a
redeployment pool and eventually terminated were compared with the
ages of 272 retained employees. Levene's test showed that the
standard deviations (9.97 and 6.94) of the age distributions of the
two groups were statistically significant. The usual
\mbox{$t$-test} found the difference of three years between the average
ages of the two groups significant (two-sided $p$-value is 0.024),
while the modified $t$-test did not (two-sided $p$-value is 0.093).
Surprisingly, the transcript of the expert testimony does not
mention any questions by the defendant about the potential implication
of the result that the age distributions of retained and laid-off
employees were similar. Comparisons showing that the termination rates of
employees aged 50 or more were higher than those of
employees under 50, however, were quite significant ($p<0.001$). This analysis
provided very strong evidence supporting the finding of age
discrimination.

\subsection{Miscellaneous Applications}

By the late 1990s researchers had documented geographical
differences in semen quality, including sperm concentration, which
raised questions about the \mbox{possible} causal roles of genetic
differences and environmental factors. Since the criteria for
recruiting study subjects, methods of laboratory analysis and
experimental design differed among the earlier studies, to eliminate
those factors as possible explanations for the basic finding, Auger
and Jouannet (\citeyear{1997Auger}) conducted a retrospective study of candidate
semen donors to sperm banks at University hospitals in eight regions
of France during the period 1973--1993. These hospitals adopted the
same guidelines for recruiting male semen donors and used similar
laboratory methods. The authors analyzed data on seminal volume,
sperm concentration, sperm count and the percentage of sperm that
were motile. As the data were not normally distributed, they made
appropriate transformations for each variable of
interest, for example, the square root transform for sperm concentration and
total sperm count. Levene's original test indicated that even the
transformed data for all four variables had statistically
significantly different variances. Hence, the authors used the Welch
analog~(\ref{6}) of ANOVA to analyze the data. The results showed
statistically significant differences among the eight regions in all
four characteristics of semen quality (all $p$-values are less than
0.0001). While these small $p$-values arose in part because the
total sample size was large (4710), varying from 226 in Caen to 1396
in Paris, the differences appear to be quite meaningful. For
instance, the mean total sperm count varied from 284 per million in
Toulouse to 409 per-million in Caen. The authors showed that these
regional differences remained statistically significant after
controlling for age, year of semen donation and number of days the
subject abstained from sex prior to sample collection.

Sexual fantasies and their content can provide insight into the
process of sexual arousal as well as gender differences in what
people find exciting. As previous research indicated that men have
more fantasies than women, Hicks and Leitenberg (\citeyear{2001Hicks}) studied
whether men and women differ in their likelihood of having sexual
fantasies about their current partner as compared to extra-dyadic
fantasies (about someone else) after controlling for the overall
difference in number of fantasies.  Using an anonymous
questionnaire, they obtained 317 surveys from students (94\%
response rate) and 273 completed surveys (24\% response rate) from faculty and
staff at a mid-sized University. Eliminating a few cases with
missing data, six outliers and 188 forms from individuals not
currently in a relationship, they analyzed 349 responses (215
females, 134 males); apparently females had a higher response rate
than males. Levene's test showed a significant gender difference in
the variance of the number of fantasies, so the Welch modified
$t$-test was used to compare the means. Men had a statistically
significantly higher number of fantasies per month than women (76.7
vs. 34.1, $t_{192} =-4.77$). To control for this gender
difference in total number of fantasies, the researchers calculated
the percentage of each respondent's fantasies that were extra-dyadic.
Since the variances of these percentages again differed by gender,
the Welch $t$-test showed that men reported a greater number of sexual
fantasies with an outsider than women (54\% vs. 36\%, $t_{311}=
-5.1$). While only a modest percentage of extra-dyadic
fantasies\break concerned former partners, on average, women had
significantly more of them than men (34\% vs. 22\%, $p=0.004$).

A regression analysis, adjusting for length of the relationship and whether one cheated
on their partner, showed that the number of prior partners a
person had was significantly more highly related to the percentage
of extra-dyadic fantasies of women than men. The percentages of fantasies
that involved someone other than their current partner was nearly identical for
men and women who had cheated on their partner (55\% vs. 53\%), implying that the major
difference between the genders in extra-dyadic fantasies occurs in
faithful partners. Since the percentages of male and female
respondents who admitted to having cheated on their current partner
were nearly identical (28\% vs. 29\%), the previous finding is not
likely to have been affected by nonresponse. For both sexes, the
percentage of fantasies that were extra-dyadic increased with the
length of the relationship. As most of the individuals in long-term
relationships were faculty and staff rather than students, the
subjects with a high degree of nonresponse, this last finding might
require further confirmation. Since the overall regression had an
$R^2$ of only 0.25, more research is needed to determine other
explanatory factors as well as improving the accuracy of the recall
data collected in similar studies.

\section{Discussion and Open Questions}\label{sec5}

Levene's original article and the statistical procedures that
developed and refined his original test enabled researchers in
many intellectual disciplines to check the validity of an important
assumption underlying the analysis of data obtained from studies
using an ANOVA design. With modern day computer programs for
calculation of statistical tests and estimators, the results in
Section \ref{sec3} show that today there is less need for a Levene-type test as a {\it
preliminary} step to decide whether a standard or Welch-modified
ANOVA test statistic should be applied, as the Welch procedure does not lose
much power when the variances are equal. With an appropriate choice for the
size of the Levene-type preliminary test, the two-stage procedure is
valid and can be reliably used in disciplines where it has become a standard
technique.

Levene's article and the subsequent literature have properly focused
users of statistics on the need to examine whether their data ``fit'' the assumptions underlying
the methods they apply. If one observes a ``borderline'' result, a
Levene-type test may be used as one of the diagnostic tools to
assess the sensitivity of the inference to potential violations of
the basic assumptions. In particular, an analog of the Sprott and
Farewell (\citeyear{1993Sprott}) use of a confidence interval for the ratio,
$\rho^2$, of both sample variances in the Behrens--Fisher problem to
assess the sensitivity of inferences on the difference of the two
means should be developed for the $k$-group setting. Using the
ratios of the mean absolute deviations from a robust estimate of the
group centers in place of the ratio of the sample variances may increase the
applicability of this technique to data from heavier tailed
distributions.

The Welch-modified $t$-test now appears in some standard textbooks
and statistical packages. Since that procedure has been shown to be
nearly as powerful as the standard one used in the equal variance
setting and has much superior control of the Type I error when the
group variances differ, authors of statistical textbooks should
consider including it in their discussion of ANOVA. The main extra
complications are the calculation of the denominator of the
statistic~(\ref{6}) and the degrees of freedom~(\ref{7}), which are
now readily carried out in statistical software. Since Levene-type
tests for equal variance or a trend in variances are easy to
describe and nearly as powerful as more complicated alternative
procedures (Pan, \citeyear{2002Pan}), these methods can now be included in statistics curriculum.

Reviewing the applied literature showed that comparing the
variability of data from several groups frequently is the scientific
question of interest. In particular, analysis of the
variability of the measurements of medical characteristics obtained from
different devices or techniques should lead to more reliable diagnosis.
Quite often the problem of interest was whether there was a decreasing or increasing trend in
the variability of the characteristic of interest that is associated
with a covariate. This was the focus of articles from a variety of fields:
the study relating characteristics of pots to the degree of
economic development, the investigations of the relationship between
the amount of information given to juries and the variability of the monetary
damages they award, or the variability of eye \mbox{examination} measurements.

The simple test described in Section~\ref{sec2}, along with related
references, should be useful to researchers concerned with similar
trend alternatives. For example, Kutner, Nachtsheim and Neter (\citeyear{2004Kutner})
describe the use of the BFL two-sample test for checking the
equality of variances of residuals from a time series regression
against a time-trend alternative. It is likely that the power of
such a test would be increased if more than two groups were formed
and the trend test was applied. Further research is needed, as the
appropriate number of groups is likely to depend on the total sample
size as well as the magnitude of the trend.

The increased power of the test will also enable researchers to use
smaller samples in those studies. Graubard and Korn (\citeyear{1987Graubard}) noted
that the choice of scores used in the Cochran--Armitage (CA) trend
test in proportions is an important topic, as they can have a noticeable effect
on the $p$-value of the test. Their point also applies to the trend test for
variances. When there are several scientifically plausible choices
for the weights, analogs of the efficiency robust methods (Zheng et al., \citeyear{2003Zheng})
developed for the CA test can be obtained, as the correlations of the
test statistics based on each set of weights can be estimated from
the data. These correlations are used in creating a suitable test statistic
that has high power over the family of scientifically plausible models of the trend.

Although there exist several methods based on Levene-type statistics
for studying differences in variability or the scale parameter of
two variables measured on paired data (Wilcox, \citeyear{1989Wilcox}; Grambsch, \citeyear{1994Grambsch}), the
visual acuity study (Rosser, Murdoch and\break Cousens, \citeyear{2004Rosser}) indicates that appropriate $k$-sample
versions should be developed. A related problem occurs when
the same technician assesses the same sample with several devices. This
topic is related to tests for the equality of variance in randomized
block designs. The survey of Schaalje and Despain (\citeyear{1996Schaale}) found that
when the block effect is mild, the method of Wilcox (\citeyear{1989Wilcox}) performs well.
When the block effect is strong and the distributions are symmetric,
a variant of Levene's test due to Yitnosumarto and O'Neill (\citeyear{1986Yitnosumarto})
is recommended. Further research is needed for the situation of asymmetric
or very heavy-tailed distributions.

Textbook discussions of ANOVA focus on comparing a relatively small
number of treatments (groups) and the large sample theory is derived
assuming that the numbers of observations in each group increase at
the same rate. In some situations the number of treatments can also
be large (Boos and Brownie, \citeyear{1995Boos}).  Bathke (\citeyear{2002Bathke},
\citeyear{2004Bathke}) examines the effect of unequal
variances in the multi-factor situation. In the commonly occurring two-factor design,
when the number of levels of the first factor, A1, increases but the number of levels
of the second, A2, remains finite, as long as the inequality in the error variances
is not related to the level of factor A1, the $F$-test for the main effect of the first
factor is almost unaffected by differences in the variances at the levels of the other factor.
The tests for the main effect of factor A2 and interaction, however, are affected.
A thorough analysis of tests of equality of variance when there are many treatments
with a modest sized sample for each one remains to be done.

In most of the applications discussed here the observations in each
group are independent random samples. It is well known (van Belle,
\citeyear{2002van}) that dependence can have a major effect on the distribution of
many standard statistics. Thus, researchers will need to design
their experiments and studies carefully to ensure that the
observations in each group are independent of each other and those
in other groups. This may not be a routine problem in studies where
the same individuals and devices are used to make the measurements.
More statistical procedures that model the dependence appropriately
and incorporate it in the analysis need to be developed.

In several large studies we reviewed there was some nonresponse or
missing data. In general, the potential effect of missing data on the
conclusions of a study should be examined, as in English, Armstrong and Kricker (\citeyear{1998Armstrong}).
In the study by Berger et al. (\citeyear{1999Berger}), only a small proportion
of data was missing, which was unlikely to affect the conclusions. Nevertheless, researchers
should be encouraged to report the pattern of missing data and any methods of imputation they
adopted in the statistical analysis.

In contrast, the probability of nonresponse in the study of
sexual fantasies (Hicks and Leitenberg, \citeyear{2001Hicks}) was highly correlated
with age, a characteristic that is related to two independent variables in the
regression predicting percentage of fantasies that were extradyadic. Thus, a
study population containing a greater proportion of older respondents might
yield different estimates of the effects of the number of prior partners and
the length of current relationship, respectively.
 Since the slope of the regression relating the proportion of extradyadic fantasies to
number of prior partners was stronger for women than for men,
whether the nonresponse rates of older males and females differed
should also be investigated. Given the recent development of
imputation and other techniques for handling missing data (Little
and Rubin, \citeyear{2002Little}; Molenberghs and Kenward, \citeyear{2007Molenberghs}), it would be useful
to explore how they can be used in these applications to
realistically assess the affect of missing data on the results of
Levene-type tests, both for homogeneity and trend.

The number of observational, rather than designed, studies we
encountered in the area of quality control or accuracy of medical
measurements indicates the importance of developing methods for
assessing the sensitivity of inferences based on tests of the
equality of variance to an unobserved variable. Hopefully, this
review will stimulate the development of methods analogous to those
used to assess the potential impact of omitted variables on the
comparison of the means or proportions from two samples (Rosenbaum,
\citeyear{2002Rosenbaum}) or in regression analysis (Dempster, \citeyear{1988Dempster}).

For cost-effectiveness many government sponsored surveys have a
complex design based on stratified multistage probability cluster
sampling, which produces estimates of population means and
proportions with larger standard errors than would be obtained from a
purely random sample of the same size (Nygard and Sandstrom, \citeyear{1989Nygard};
Korn and Graubard, \citeyear{1999Korn}). Appropriate modifications of Levene-type
tests for variance or measures of relative variability should be
useful when the status of several sub-groups of the population is
studied.

\section*{Acknowledgments}
The research of Professor Gastwirth was supported in part by NSF Grant
SES-0317956.
The research of
Professor Gel was in part supported by a Grant from NSERC of
Canada and was made possible by
the facilities of SHARCNET.

\end{document}